\documentclass[12pt]{article}
\usepackage{a4wide}
\usepackage{amssymb}
\usepackage{amsmath}
\usepackage{graphicx}
\usepackage{slashed}
\usepackage{verbatim}

\def\XXint#1#2#3{{\setbox0=\hbox{$#1{#2#3}{\int}$}
     \vcenter{\hbox{$#2#3$}}\kern-.52\wd0}}

\newcommand{\be}{\begin{equation}}\newcommand{\ee}{\end{equation}}
\newcommand{\bea}{\begin{eqnarray}} \newcommand{\eea}{\end{eqnarray}}

\def\makeatletter{\catcode`\@=11}
\makeatletter
\def\mathbox#1{\hbox{$\m@th#1$}}%
\def\math@ccstyles#1#2#3#4#5#6#7{{\leavevmode
      \setbox0\mathbox{#6#7}%
      \setbox2\mathbox{#4#5}%
      \dimen@ #3%
      \baselineskip\z@\lineskiplimit#1\lineskip\z@
      \vbox{\ialign{##\crcr
             \hfil \kern #2\box2 \hfil\crcr
             \noalign{\kern\dimen@}%
             \hfil\box0\hfil\crcr}}}}
\def\mathaccstyles{\math@ccstyles\maxdimen}
\def\maththroughstyles{\math@ccstyles{-\maxdimen}}
\def\unity%
 {\maththroughstyles{.45\ht0}\z@\displaystyle {\mathchar"006C}\displaystyle 1}

\makeatletter \@addtoreset{equation}{section} \makeatother

\begin{document}

\setcounter{table}{0}

\begin{flushright}\footnotesize

\vspace{0.6cm}
\end{flushright}

\mbox{}
\vspace{0truecm}
\linespread{1.1}

\centerline{\LARGE \bf Operator Mixing in Large $N$}
\bigskip
\centerline{\LARGE \bf   Superconformal Field Theories on  $\mathbb{S}^4$  }
\bigskip
\centerline{\LARGE \bf   and Correlators with Wilson loops}

\vspace{.5cm}

 \centerline{\LARGE \bf }

\vspace{1.5truecm}

\centerline{
    {\large \bf Diego Rodriguez-Gomez${}^{a}$} \footnote{d.rodriguez.gomez@uniovi.es}
    {\bf and}
    {\large \bf Jorge G. Russo ${}^{b,c}$} \footnote{jorge.russo@icrea.cat}}

\vspace{1cm}
\centerline{{\it ${}^a$ Department of Physics, Universidad de Oviedo}} \centerline{{\it Avda.~Calvo Sotelo 18, 33007  Oviedo, Spain}}
\medskip
\centerline{{\it ${}^b$ Instituci\'o Catalana de Recerca i Estudis Avan\c{c}ats (ICREA)}} \centerline{{\it Pg.Lluis Compayns, 23, 08010 Barcelona, Spain}}
\medskip
\centerline{{\it ${}^c$ Departament de F\' \i sica Cu\' antica i Astrof\'\i sica and Institut de Ci\`encies del Cosmos}} \centerline{{\it Universitat de Barcelona, Mart\'i Franqu\`es, 1, 08028
Barcelona, Spain }}
\vspace{1cm}

\centerline{\bf ABSTRACT}
\medskip
We find a general formula for the operator mixing on the $\mathbb{S}^4$ of chiral primary operators (CPO) for the
${\cal N}=4$ theory at large $N$ in terms of Chebyshev polynomials.
As an application, we compute the correlator of a CPO and a Wilson loop,
reproducing an earlier result 
by Giombi and Pestun obtained from a two-matrix model proposal.
Finally, we discuss  a simple method to obtain correlators 
in general ${\cal N}=2$ superconformal field theories in perturbation theory 
in terms of correlators of the ${\cal N}=4$ theory.

\noindent

\newpage

\section{Introduction}

In ${\cal N}=2$ superconformal field theories (SCFT's), correlators of a special type of scalar operators known as chiral primaries (CPO's) can be computed exactly from a deformed matrix model obtained by adding suitable source terms to the theory on $\mathbb{S}^4$ \cite{Gerchkovitz:2016gxx}. Since the construction involves a conformal map from  $\mathbb{S}^4$ into $\mathbb{R}^4$, and
 because the regulated theory on  $\mathbb{S}^4$ breaks the $U(1)_R$ symmetry,  mixings among operators with different R-symmetries are possible. Alternatively, since the radius of the  $\mathbb{S}^4$ sets a scale which can soak up dimensions, operators of different dimensions can mix. These mixings can be viewed as a manifestation of the conformal anomaly.

In order to take care of the operator mixings, Gerchkovitz {\it et al.} \cite{Gerchkovitz:2016gxx} proposed a Gram-Schmidt procedure to disentangle the correct operators corresponding to those in $\mathbb{R}^4$. Such procedure was carried out explicitly in \cite{Rodriguez-Gomez:2016ijh} for $\mathcal{N}=4$ Super Yang-Mills theory (SYM) in the large $N$ limit, where the structure of the CPO's dramatically simplifies as only single-trace operators are kept. In particular, it was shown how the expected free-field result is recovered. This is a consequence of a celebrated non-renormalization theorem \cite{Lee:1998bxa,Freedman:1998tz,D'Hoker:1998tz,Penati:1999ba,Baggio:2012rr}.

As discussed in \cite{Gerchkovitz:2016gxx}, the coefficients encoding the mixings among operators are subject to an  ambiguity; as they are defined modulo the addition of holomorphic and antiholomorphic functions of the couplings.
 This ambiguity can nevertheless be removed by taking appropriate derivatives. Thus, the resulting differentiated mixing coefficients should be observables, very plausibly linked to new manifestations of the conformal anomaly. It is therefore natural to expect the structure of the operator mixings (modulo the holomorphic ambiguity) to be very interesting. 

In this note we explore the structure of the operator mixing, focusing in the case of $\mathcal{N}=4$ SYM in the large $N$ limit (we provide some results  for $\mathcal{N}=2$ superconformal QCD as well). As we will see, the mixing coefficients nicely fit into a pattern corresponding to the Chebyshev polynomials (see eq. \eqref{Tns}).
We will show that this form is exactly what is needed in order to recover the expected correlation functions from the matrix model  \cite{Rodriguez-Gomez:2016ijh}. Moreover, these operator mixings in the form of Chebyshev polynomials are also exactly what is needed to recover the expected form of the correlators between CPO's and Wilson loops   \cite{Giombi:2009ds,Giombi:2012ep,Bassetto:2009rt,Bassetto:2009ms,Bonini:2014vta}, thus providing a highly non-trivial check of the relevance of the mixing structure.

It is worth emphasizing that the correlator between a CPO and a Wilson loop has been previously computed in   \cite{Giombi:2009ds,Giombi:2012ep,Bassetto:2009rt,Bassetto:2009ms,Bonini:2014vta}. Since the propagators evaluated on the loop become position independent, the computation can be done through a certain two-matrix model encoding the combinatorics of the Feynman diagrams. Interestingly, such matrix model is related to 2d Yang-Mills theory \cite{Pestun:2007rz}. Here we shall arrive at the same result through a completely different path, thus suggesting an intriguing large $N$ ``duality" between matrix models.

This paper is organized as follows: in section \ref{OperatorMixing} we briefly review the construction of  \cite{Gerchkovitz:2016gxx} to compute correlators of CPO's and explain the relevance of the operator mixing. We then specialize to $\mathcal{N}=4$ SYM and 
establish the generic expression \eqref{Tns} encoding such operator mixing.
In section \ref{WilsonCPO} we turn to the computation of the correlator of a Wilson loop and a CPO. Following the same logic,
 we uncover a new representation for these correlators in terms of the  $\mathbb{S}^4$ matrix model with the insertion of operators in the form of the Chebyshev polynomials, as in eq. \eqref{Tns}. This  reproduces the same result obtained in \cite{Giombi:2009ds} in a highly non-trivial way,
and confirms the general form of the operator mixing on  $\mathbb{S}^4$. In section \ref{SQCD} we turn to $\mathcal{N}=2$ conformal SQCD, for which we compute the analogous correlator among a Wilson loop and a CPO. Along the way, we find that correlators of CPO's in $\mathcal{N}=2$ superconformal QCD can be written in terms of correlators of $\mathcal{N}=4$ CPO's and prove a conjecture \cite{Rodriguez-Gomez:2016ijh} about the 2-point function for CPO's in $\mathcal{N}=2$ SQCD with arbitrary
  $SU(N)$ gauge group. We close in section \ref{conclusions} with concluding remarks.

\section{Operator mixing in  $\mathbb{S}^4$ from correlators of \\ chiral primary operators}\label{OperatorMixing}

Superconformal primary operators, annihilated by the superconformal supersymmetry generators $S_{\alpha}^i$ and $\bar{S}_{\dot{\alpha}}^i$, play a prominent role in $\mathcal{N}=2$ theories. Among those, chiral operators (CPO), annihilated as well by the Poincar\' e supercharges of one chirality $\bar{Q}_{\dot{\alpha}}^i$, are of special relevance. In particular, strong arguments suggest that these operators are always Lorentz scalars satisfying the BPS bound $\Delta=\frac{R}{2}$, being $R$ the $U(1)_R$ charge. This implies that their OPE is non-singular with constant structure functions, endowing these operators with a ring structure. Very recently, there has been much progress in understanding the structure of such ring \cite{Papadodimas:2009eu,Baggio:2014ioa,Baggio:2015vxa,Gerchkovitz:2014gta,Gomis:2014woa,Gerchkovitz:2016gxx,Rodriguez-Gomez:2016ijh}. In particular, in \cite{Gerchkovitz:2016gxx} it was argued that correlators of CPO's in $\mathbb{R}^4$ can be extracted from the partition function of a deformed version of the theory on  $\mathbb{S}^4$. This is possible due to a Ward identity on the  $\mathbb{S}^4$, which relates  the integrated correlator of particular combination including the top component of the $\mathcal{N}=2$  chiral superfield where the CPO lives, denoted by $\mathcal{C}$, with the (unintegrated) correlator of CPO's $\mathcal{O}$ at North and South poles of the sphere (see \cite{Gerchkovitz:2014gta,Gomis:2014woa,Gerchkovitz:2016gxx} for more details) as

\begin{equation}
\label{Ward}
\Big(\frac{1}{32\pi^2}\Big)^2\int_ {\mathbb{S}^4} d^4x \sqrt{g(x)}\int_ {\mathbb{S}^4} d^4y\sqrt{g(y)}\,\langle \mathcal{C}_n(x)\overline{\mathcal{C}}_m(y)\rangle_ {\mathbb{S}^4}=\langle \mathcal{O}_n(N)\overline{O}_m(S)\rangle_ {\mathbb{S}^4}\, .
\end{equation}
Furthermore,  due to supersymmetry, it holds that correlators on $\mathbb{R}^4$ satisfy that 
\cite{Baggio:2014ioa}
$$
\langle \mathcal{O}_{n_1}(x_1)\cdots \mathcal{O}_{n_n}(x_n)\overline{O'}_{m}(y) \rangle_{\mathbb{R}^4}=\langle O_n(x_1)\overline{O'}_{m}(y)\rangle_{\mathbb{R}^4}\ ,
$$ 
for $O_n(x)\equiv \mathcal{O}_{n_1}(x)\cdots \mathcal{O}_{n_n}(x)$. 
Since these operators are CPO's and thus satisfy a BPS bound, $R$-charge conservation sets  their correlator to zero unless $\Delta_O=\Delta_{O'}$. Thus, on general grounds their correlator is

\begin{equation}
\langle O_n(x)\overline{O'}_m(0)\rangle_{\mathbb{R}^4} = \frac{C_{nm}}{|x|^{2\Delta_O}}\delta_{\Delta_O\Delta_{O'}}\, .
\end{equation}
This can now be recast as 

\begin{equation}
|x|^{2\Delta_O}\, \langle O_n(x)\overline{O'}_m(0)\rangle_{\mathbb{R}^4} = C_{nm}\delta_{\Delta_O\Delta_{O'}}\, .
\end{equation}
One can now observe that 

\begin{equation}
\lim_{x\rightarrow \infty} |x|^{2\Delta_O} O(x)=4^{\Delta_O} \lim_{x\rightarrow \infty} (1+\frac{\vec{x}^2}{4})^{\Delta_O} O(x)\, .
\end{equation}
Since $\mathbb{R}^4$ and  $\mathbb{S}^4$ are related by a conformal transformation as

\begin{equation}
\label{confo}
ds^2_{\mathbb{R}^4}= \left(1+\frac{\vec{x}^2}{4}\right)^2\,ds^2_ {\mathbb{S}^4}\, ,
\end{equation}
and, given that $x\rightarrow \infty$ in the  $\mathbb{S}^4$ corresponds to the North pole, one has $\lim_{x\rightarrow \infty} |x|^{2\Delta_O} O(x)=4^{\Delta_O}O(N)$, where we used the conformal map induced by (\ref{confo}).
Conversely, since $x\rightarrow 0$ corresponds to the South pole, $O(0)=\lim_{x\rightarrow 0} (1+\frac{\vec{x}^2}{4})^{\Delta} O(x)=O(S)$. Thus, this allows one to relate the $\mathbb{R}^4$ correlator to the  $\mathbb{S}^4$ correlator as

\begin{equation}
 \langle\Big( \lim _{x\rightarrow \infty} |x|^{2\Delta_O} O_n(x)\Big)\overline{O'}_m(0)\rangle_{\mathbb{R}^4} = 4^{\Delta_O} \,\langle O_n(N)\overline{O'}_m(S)\rangle_ {\mathbb{S}^4} = C_{nm}\delta_{\Delta_O\Delta_{O'}}\, .
\end{equation}
Therefore, combining this with \eqref{Ward}, we can write 

\begin{equation}
\Big(\frac{1}{32\pi^2}\Big)^2\int_ {\mathbb{S}^4} d^4x \sqrt{g(x)}\int_ {\mathbb{S}^4} d^4y\sqrt{g(y)}\,\langle \mathcal{C}_n(x)\overline{\mathcal{C}}_m(y)\rangle_ {\mathbb{S}^4}=4^{-\Delta} \,C_{nm}\, .
\end{equation}

One can  consider a deformed version of the  $\mathbb{S}^4$ partition function by adding a source $\tau_m$ for all CPO's, obtaining in this way a deformed partition function $\mathcal{Z}[\{\tau_m,\,\bar{\tau}_m\}]$ \cite{Gerchkovitz:2016gxx}. Here
$\tau_2=\tau_{\rm YM}$, where, as usual, ${\rm Im}\tau_{\rm YM}=\frac{4\pi}{g_{\rm YM}^2}$. 
Correlators can then be computed by 
\begin{equation}
\frac{1}{\mathcal{Z}[\{\tau_m,\,\bar{\tau}_m\}]}\partial_{\tau_m}\partial_{\bar{\tau}_m}\mathcal{Z}[\{\tau_m,\,\bar{\tau}_m\}]=\Big(\frac{1}{32\pi^2}\Big)^2\int_ {\mathbb{S}^4} d^4x \sqrt{g(x)}\int_ {\mathbb{S}^4} d^4y\sqrt{g(y)}\,\langle \mathcal{C}_n(x)\overline{\mathcal{C}}_m(y)\rangle_ {\mathbb{S}^4}\ .
\end{equation}

There is, however, one  important subtlety, namely that, due to the conformal anomaly, on the  $\mathbb{S}^4$ there is a highly non-trivial operator mixing. This is expected, since the  $\mathbb{S}^4$ theory preserves the supergroup $osp(2|4)$, which contains the $SU(2)_R$ symmetry but breaks the $U(1)_R$ symmetry. Thus, mixings among different CPO's are possible. Indeed, denoting by $r$ the radius of  the $\mathbb{S}^4$, a given operator $O_{\Delta}$ of dimension $\Delta$ on $\mathbb{R}^4$, when mapped into  $\mathbb{S}^4$, generically mixes with all operators with lower dimensions in steps of 2, that is

\begin{equation}
\label{map}
O_{\Delta}^{\mathbb{R}^4}\rightarrow O_{\Delta}^ {\mathbb{S}^4}+\alpha_{\Delta}^{(2)}\frac{1}{r^2}\,O_{\Delta-2}^ {\mathbb{S}^4}+\alpha_{\Delta}^{(4)} \frac{1}{r^4}\,O_{\Delta-4}^ {\mathbb{S}^4}+\cdots \, .
\end{equation}
Due to this effect, in order to find the expected $R$-charge conservation on $\mathbb{R}^4$, when mapping  the  $\mathbb{S}^4$ computation back into the $\mathbb{R}^4$, the operator mixing must be disentangled. This can be accomplished by a Gram-Schmidt orthogonalization procedure.
From now on, we set $r=1$.

Following \cite{Gerchkovitz:2016gxx}, in \cite{Rodriguez-Gomez:2016ijh} correlators of CPO's in $\mathcal{N}=4$ SYM have been computed solving the deformed matrix model in the large $N$ limit. In this limit, the set of CPO's dramatically simplifies as only single-trace operators contribute. Thus, the operators in the chiral ring are of the form $O_n^{\mathbb{R}^4}={\rm Tr}\phi^n$, being $\phi$ one of the complex scalars in the theory (that interpreted to live in the vector multiplet when the theory is viewed as a $\mathcal{N}=2$ theory). Upon mapping into the  $\mathbb{S}^4$ the operators ${\rm Tr}\phi^n|_ {\mathbb{S}^4}$ acquire a VEV due to the mixing with the identity given by \cite{Rodriguez-Gomez:2016ijh}\footnote{This VEV can also be read from the VEV of the 1/2 BPS circular Wilson loop, which at large $N$ is given by
\cite{Erickson:2000af,Drukker:2000rr} $\langle W\rangle =\langle {\rm Tr} \exp(2\pi\phi) \rangle =\frac{2}{\sqrt{\lambda}}
I_1(\sqrt{\lambda})$, by expanding the exponential (see (\ref{wka}), (\ref{W}) below).}

\begin{equation}
\label{VEV}
\langle {\rm Tr}\phi^{2r}\rangle_ {\mathbb{S}^4}=N\Big(\frac{\lambda}{4\pi^2}\Big)^r\frac{\Gamma(r+\frac{1}{2})}{\sqrt{\pi}\,(r+1)!} = N\Big(\frac{\lambda}{4\pi^2}\Big)^r\frac{(2r)!}{4^r r! (r+1)!}\ ,
\end{equation}
and $\langle {\rm Tr}\phi^{2r+1}\rangle_ {\mathbb{S}^4}=0$
It is  useful to define the operator $O_n^ {\mathbb{S}^4}$ as the VEV-less version of ${\rm Tr}\phi^n$ in  $\mathbb{S}^4$, \textit{i.e.} 
\be
\label{sinvev}
O^ {\mathbb{S}^4}_n={\rm Tr}\phi^n|_ {\mathbb{S}^4}-\langle {\rm Tr}\phi^{n}\rangle_ {\mathbb{S}^4}\unity \ .
\ee
 Then, the large-$N$ correlators of $O_n^ {\mathbb{S}^4}$ are given by \cite{Rodriguez-Gomez:2016ijh}

\begin{eqnarray}
\label{S4correlators}
\langle O^ {\mathbb{S}^4}_{2n}  \overline{O}^ {\mathbb{S}^4}_{2r}\rangle_ {\mathbb{S}^4} &=&\Big(\frac{\lambda}{4\pi^2}\Big)^{n+r} \frac{\Gamma(n+\frac{1}{2})\Gamma(r+\frac{1}{2})}{\pi\,(n+r)\Gamma(n)\Gamma(r)}\, ; \\ 
\langle O^ {\mathbb{S}^4}_{2n+1}  \overline{O}^ {\mathbb{S}^4}_{2r+1}\rangle_ {\mathbb{S}^4} &=&\Big(\frac{\lambda}{4\pi^2}\Big)^{n+r+1} \frac{\Gamma(n+\frac{3}{2})\Gamma(r+\frac{3}{2})}{\pi\,(n+r+1)\Gamma(n+1)\Gamma(r+1)}\, .
\end{eqnarray}

The map \eqref{map} between $O_n^{\mathbb{R}^4}$ and $O_n^ {\mathbb{S}^4}$ arising from the Gram-Schimidt procedure
was explicitly calculated  in \cite{Rodriguez-Gomez:2016ijh} for the first few operators. One finds 
\begin{eqnarray}
\label{OperatorsN=4}
O_1^{\mathbb{R}^4}&=&O^ {\mathbb{S}^4}_1\,; \nonumber \\
O_2^{\mathbb{R}^4}&=&O^ {\mathbb{S}^4}_2\,;\nonumber \\
O_3^{\mathbb{R}^4}&=& O^ {\mathbb{S}^4}_3-3\frac{\lambda}{(4\pi)^2}O^ {\mathbb{S}^4}_1\,; \nonumber\\
O_4^{\mathbb{R}^4}&=&O^ {\mathbb{S}^4}_4-4\frac{\lambda}{(4\pi)^2} O^ {\mathbb{S}^4}_2\,;\nonumber \\
O_5^{\mathbb{R}^4}&=&O^ {\mathbb{S}^4}_5-5\frac{\lambda}{(4\pi)^2}O^ {\mathbb{S}^4}_3+5\frac{\lambda^2}{(4\pi)^4}O^ {\mathbb{S}^4}_1\,;\nonumber \\
O_6^{\mathbb{R}^4}&=&O^ {\mathbb{S}^4}_6-6\frac{\lambda}{(4\pi)^2} O^ {\mathbb{S}^4}_4+9\frac{\lambda^2}{(4\pi)^4}O^ {\mathbb{S}^4}_2\,;\nonumber \\
&&\cdots
\end{eqnarray}
With these ingredients, it was shown in \cite{Rodriguez-Gomez:2016ijh} that the first few correlators of $O_n^{\mathbb{R}^4}$ exactly match the general formula

\begin{equation}
\label{correlator}
\langle O_n^{\mathbb{R}^4}(x)\overline{O}_m^{\mathbb{R}^4}(0)\rangle_{\mathbb{R}^4}=\frac{\delta_{nm}}{|x|^{2\Delta_n}}\,\frac{\Delta_n\,\lambda^{\Delta_n}}{(2\pi)^{2\Delta_n}}\,,
\end{equation} 
being $\lambda$ the 't Hooft coupling. This formula coincides with the free field theory result due to a 
non-renormalization theorem   \cite{Lee:1998bxa,Freedman:1998tz,D'Hoker:1998tz,Penati:1999ba,Baggio:2012rr}.

\subsection{A general expression for the operator mixing\\ in large $N$  $\mathcal{N}=4$ SYM}

Upon inspection of \eqref{OperatorsN=4} one can identify the general pattern underlying the mixing structure. We find that the operator mixing \eqref{OperatorsN=4} is encoded in the general expression

\begin{equation}
\label{TnsSYMB}
O_n^{\mathbb{R}^4}\rightarrow 2 \Big(\frac{\lambda}{(4\pi)^2}\Big)^{\frac{n}{2}}\,\Big[T_n(\frac{2\pi x}{\sqrt{\lambda}})-T_n(0)\Big]\, ,
\end{equation}
where  $T_n(x)$ is the $n$-th Chebyshev polynomial
and it is understood that the term $x^k$ in the polynomial stands for the VEV-less operator $O_k^ {\mathbb{S}^4}$, defined in (\ref{sinvev}).
The formula can be given in a more explicit form as

\bea
\label{Tns}
&& O_n^{\mathbb{R}^4} = 2 \Big(\frac{\lambda}{(4\pi)^2}\Big)^{\frac{n}{2}}\,  {\rm Tr} \Big[T_n \big(\frac{2\pi}{\sqrt{\lambda }} \ \phi \big)\Big]\, ,\ \qquad n\neq 2\ ,
\nonumber\\
&& O_2^{\mathbb{R}^4} =  \frac{\lambda}{(4\pi)^2}\, \left(2  {\rm Tr} \Big[T_2 \big(\frac{2\pi}{\sqrt{\lambda }} \ \phi \big)\Big] +1\right)\, ;
\eea
where we have used the property
\be
\langle  {\rm Tr} \Big[T_n \big(\frac{2\pi}{\sqrt{\lambda }} \ \phi \big)\Big]\rangle= -\frac{\delta_{2,n}}{2} \ .
\ee
This can be proved by substituting the expressions for the VEV given in (\ref{VEV}).\footnote{The particular combination of factors $ 2 b^{\frac{n}{2}} T_n(\frac{x}{2b})$ goes in the literature under the name of bivariate Chebyshev polynomials.}

Using \eqref{Tns} and the sphere correlators \eqref{S4correlators}, we can now prove  in full generality \eqref{correlator}. To that matter, note that

\begin{equation}
\label{extse}
T_n(x)=\frac{n}{2} \sum_{k=0}^{\lfloor \frac{n}{2}\rfloor} (-1)^k\,\frac{(n-k-1)!}{k!(n-2k)!} (2x)^{n-2k}\, .
\end{equation}
For even $n$, $T_n$ only contains even powers of $x$, while for odd $n$, $T_n$ contains only odd powers of $x$. 
Since the correlators \eqref{S4correlators} vanish for $\langle O_{2n}^ {\mathbb{S}^4}\overline{O}^ {\mathbb{S}^4}_{2m+1}\rangle$, it follows that $\langle O_{2n}^{\mathbb{R}^4}\overline{O}^{\mathbb{R}^4}_{2m+1}\rangle =0$. 

Consider now the case of $\langle O_{2n+1}^{\mathbb{R}^4}\overline{O}^{\mathbb{R}^4}_{2m+1}\rangle_{\mathbb{R}^4}$. 
Substituting the  Chebyshev polynomials (\ref{extse}), we find

\begin{eqnarray}
\langle O_{2n+1}^{\mathbb{R}^4}\overline{O}^{\mathbb{R}^4}_{2m+1}\rangle_{\mathbb{R}^4}
&=&4^{2n+1} (2n+1)(2m+1) \times  \\ \nonumber  && \sum_{k=0}^n \sum_{q=0}^m \frac{ (-1)^{k+q}\,g^{k+q}\, (2n-k)!\,(2m-q)!}{k!\,q!\,(2n-2k+1)!\,(2m-2q+1)!}\,\langle O_{2(n-k)+1}^ {\mathbb{S}^4}\overline{O}_{2(m-q)+1}^ {\mathbb{S}^4}\rangle_ {\mathbb{S}^4}\, .
\end{eqnarray}
where we have set, for simplicity, $\frac{\lambda}{(4\pi)^2}\equiv g$. The overall numerical factor $4^{2n+1}$ stands for the $4^{\Delta}$ factor discussed above, originating from the conformal map ${\mathbb S}^4\to {\mathbb R}^4$. 
Substituting here the formula \eqref{S4correlators},
we obtain
\begin{eqnarray}
\langle O_{2n+1}^{\mathbb{R}^4}\overline{O}^{\mathbb{R}^4}_{2m+1}\rangle_{\mathbb{R}^4}
&=&4^{2n+1} (2n+1)(2m+1) g^{n+m+1}\times  \\ \nonumber  && \sum_{k=0}^n \sum_{q=0}^m \frac{(-1)^{k+q}}{k!\,q! }\, \frac{ (2n-k)!\,(2m-q)!\ }{(n+m-k-q+1)(n-k)!^2(m-q)!^2}
\nonumber\\
&=&  (2n+1) \left( \frac{\lambda}{(2\pi)^2}\right)^{2n+1}\delta_{n,m}\, .
\end{eqnarray}
Strikingly,  this  exactly reproduces \eqref{correlator} for odd-dimensional operators.

Next, for  even-dimensional operators the analogous computation reads
\begin{eqnarray}
\langle O_{2n}^{\mathbb{R}^4}\overline{O}^{\mathbb{R}^4}_{2m}\rangle_{\mathbb{R}^4}
&=&4^{2n} (2n)(2m) g^{n+m}\times  \\ \nonumber  && \sum_{k=0}^{n-1} \sum_{q=0}^{m-1}  \frac{(-1)^{k+q}\ (2n-k-1)!\,(2m-q-1)! }{(n+m-k-q)k!\,q! (n-k)!(n-k-1)! (m-q)!(m-q-1)!}
\nonumber\\
&=& \ 2n \left( \frac{\lambda}{(2\pi)^2}\right)^{2n}\   \delta_{n,m} \ ,
\end{eqnarray}
which exactly reproduces the general result \eqref{correlator} for even-dimensional operators. 

We have seen that the mapping between $\mathbb{R}^4$ operators and  $\mathbb{S}^4$ operators in large $N$ $\mathcal{N}=4$ SYM is given by the Chebyshev polynomials \eqref{Tns}. We can now extract the explicit formulas for the operator mixing coefficients. Comparing \eqref{map} with \eqref{Tns}, we have that

\begin{equation}
 \alpha_n^{(2k)}=(-1)^k\,n\, \frac{(n-k-1)!}{k!(n- 2k)!} \Big(\frac{\lambda}{(4\pi)^2}\Big)^{k}\, .
\end{equation}
As discussed in \cite{Gerchkovitz:2016gxx}, these coefficients are scheme-dependent and can be redefined as 
$$
\alpha_n^{(2k)}\rightarrow \alpha_n^{(2k)}+f_n^{(2k)}(\tau_{YM})+\bar{f}_n^{(2k)}(\bar{\tau}_{YM})\ .
$$ 
Nevertheless, an unambiguous quantity, which we will denote as $A_n^{(2k)}$, can be constructed as

\begin{equation}
A_n^{(2k)}=\partial_{\tau_{YM}}\partial_{\bar{\tau}_{YM}}\alpha_n^{(2k)}\, .
\end{equation}
In the case at hand, we find

\begin{equation}
\label{aann}
A_n^{(2k)}= (-1)^{k} \frac{n(k+1)(n-k-1)!}{4N^2 (k-1)! (n-2k)!}\frac{\lambda^{2+k}}{(4\pi)^{2+2k}}\, .
\end{equation}
It would be extremely interesting to understand the physical significance of the $A_n^{(2k)}$, in particular,
as a manifestation of the conformal anomaly. 

\section{Correlators of chiral primary operators and Wilson loops in $\mathcal{N}=4$ SYM}\label{WilsonCPO}

Let us turn our attention to the correlator between a chiral primary operator and a Wilson loop \cite{Giombi:2009ds,Giombi:2012ep}.  We consider the family of 1/8 supersymmetric
Wilson loops introduced in \cite{Drukker:2007qr,Drukker:2007dw,Drukker:2007yx}. These are constrained to live in an $S^2$ inside $\mathbb{R}^4$ at $x_4=0$ and satisfying $x_1^2+x_2^2+x_3^2=r^2$. Generically, these Wilson loops couple to three out of the six real scalars of the $\mathcal{N}=4$ SYM theory, which we may  denote by $X_i$, as

\begin{equation}
\label{W}
W_R(C)={\rm Tr}_R {\rm P}\,e^{\int_C (A_j+i \epsilon_{ijk} X_i\,\frac{y^k}{r})dy^j}\,.
\end{equation}

\noindent A specially interesting class of operators are those of the form ${\rm Tr}(X_{\rm n}+iY)^n$, where $X_{\rm n}=\sum\frac{x_i}{r}X_i$ and  $Y$ stands for any of the remaining  three scalars \cite{Drukker:2009sf,Giombi:2009ds}. These operators share 2 supersymmetries with the Wilson loop. 

We may now choose the loop to lie on the maximal circle --so that it becomes 1/2 BPS-- at $x^3=0$, when it couples only to $X_3$. Moreover, we can insert the scalar operators at $\{x_1,\,x_2,\, x_3\}=\{ \pm r,\, 0,\, 0\} $, so that $X_{\rm n}=\pm X_3$. Then, the combination $\phi=\pm X_3+i Y$ is just a CPO 
 for which the results of \cite{Gerchkovitz:2016gxx,Rodriguez-Gomez:2016ijh} apply. Thus, let us consider the correlator $\langle O_n^{\mathbb{R}^4}(0) W\rangle_{\mathbb{R}^4}$, with $O^{\mathbb{R}^4}_n={\rm Tr}\phi^n(0)$. 
Note that $r$ is then  the distance between the insertion point and the center of the loop. 
Following the same procedure as in the previous section, we can then map this into the  $\mathbb{S}^4$, so that

\begin{equation}
\label{WPhi}
\langle O_n^{\mathbb{R}^4}(0) W\rangle_{\mathbb{R}^4}=\frac{1}{r^n}\, \langle O_n^ {\mathbb{R}^4}(S) W\rangle_ {\mathbb{S}^4}\, .
\end{equation}

In the remainder of this section, we will compute \eqref{WPhi} and then compare it against the results in \cite{Giombi:2009ds}. To that matter, the map \eqref{Tns} will be crucial.

It is useful to first compute the vacuum expectation value of the 1/2 BPS Wilson loop,  which amounts to the insertion  of ${\rm Tr}e^{2\pi\phi}$.
Expanding the exponential, it follows that

\begin{equation}\label{wka}
\langle W\rangle_ {\mathbb{S}^4} = \sum_{k=0}^{\infty}\frac{(2\pi)^n}{n!} \langle {\rm Tr}\phi^n\rangle\ .
\end{equation}
Substituting \eqref{VEV} into \eqref{wka}, we find
\begin{equation}
\label{W}
\langle W\rangle_ {\mathbb{S}^4} = \frac{2}{\sqrt{\lambda }}\ I_1\big( \sqrt{\lambda } \big)\ ,
\end{equation}
where $I_n$ denotes, as usual, the modified Bessel function of the first kind.
This reproduces
the familiar formula for the VEV of the circular Wilson loop found in
in \cite{Erickson:2000af,Drukker:2000rr}.

Next, consider
\begin{equation}
\langle O_n^{\mathbb{R}^4}W\rangle_ {\mathbb{S}^4} = \sum_{r=0}^{\infty}\frac{(2\pi)^r}{r!} \langle O_n^{\mathbb{R}^4} {\rm Tr} \overline{\phi}^r\rangle_ {\mathbb{S}^4}\ .
\end{equation}
Here $O_n^{\mathbb{R}^4}$ is to be interpreted through the operator mixing formula \eqref{Tns}, that is, in terms of the VEV-less operators in  $\mathbb{S}^4$ inside the $n$-th Chebyshev polynomial. Note that ${\rm Tr}\overline{\phi}^r$ does not correspond to a VEV-less operator, and hence the expressions \eqref{S4correlators} do not directly apply. Nevertheless, we may write $\langle O_n^{\mathbb{R}^4} {\rm Tr} \overline{\phi}^r\rangle_ {\mathbb{S}^4}=\langle O_n^{\mathbb{R}^4} O_r^ {\mathbb{S}^4}\rangle_ {\mathbb{S}^4}+\langle {\rm Tr}\overline{\phi}^r\rangle_ {\mathbb{S}^4}\langle O_n^{\mathbb{R}^4} \rangle_ {\mathbb{S}^4}$. Since $O_n^{\mathbb{R}^4}$ is a polynomial in terms of the $O_n^ {\mathbb{S}^4}$, whose VEV vanishes, the last term is zero. Therefore we can write

\begin{equation}
\langle O_n^{\mathbb{R}^4}W\rangle_ {\mathbb{S}^4} = \sum_{r=0}^{\infty}\frac{(2\pi)^r}{r!} \langle O_n^{\mathbb{R}^4} \overline{O}_r^ {\mathbb{S}^4}\rangle_ {\mathbb{S}^4}\ .
\end{equation}

\noindent Writing the operator  $O_n^{\mathbb{R}^4}$ in terms of the Chebyshev polynomials,  \eqref{Tns},
using the expansion \eqref{extse} and  computing the two-point functions using \eqref{S4correlators}, we find  
\bea
\langle O_{2n}^{\mathbb{R}^4}W\rangle_ {\mathbb{S}^4} 
&=&
2n\sum_{s=0}^\infty \frac{(2\pi)^{2s}}{s! (s-1)!}  \Big(\frac{\lambda}{(4\pi)^2}\Big)^{n+s} \sum_{k=0}^{n-1} \frac{(-1)^k (2n-k-1)!}{(n+s-k) k! (n-k-1)!(n-k)!}
\nonumber\\
&=&
2n\sum_{s=n}^\infty \frac{(2\pi)^{2s}}{(s-n)! (s+n)!}  \Big(\frac{\lambda}{(4\pi)^2}\Big)^{n+s} 
\nonumber\\
&=&
2n \left(\frac{\lambda}{8\pi}\right)^{2n}\sum_{r=0}^\infty \frac{1}{r! (r+2n)!} \Big(\frac{\lambda}{4}\Big)^{r} 
\nonumber\\
&=& 2n\,\Big(\frac{\lambda}{(4\pi)^2}\Big)^{n}\,I_{2n}(\sqrt{\lambda})\, ,
\eea
Similarly,
\bea
\langle O_{2n+1}^{\mathbb{R}^4}W\rangle_ {\mathbb{S}^4} 
&=&
(2n+1)\sum_{s=0}^\infty \frac{(2\pi)^{2s+1}}{s!^2}  \Big(\frac{\lambda}{(4\pi)^2}\Big)^{n+s+1} \sum_{k=0}^{n} \frac{(-1)^k (2n-k)!}{(n+s-k+1) k! (n-k)!^2}
\nonumber\\
&=&
(2n+1)\sum_{s=n}^\infty \frac{(2\pi)^{2s+1}}{(s-n)! (s+n+1)!}  \Big(\frac{\lambda}{(4\pi)^2}\Big)^{n+s+1} 
\nonumber\\
&=&
(2n+1) \left(\frac{\lambda}{8\pi}\right)^{2n+1}\sum_{r=0}^\infty \frac{1}{r! (r+2n+1)!} \Big(\frac{\lambda}{4}\Big)^{r} 
\nonumber\\
&=& (2n+1)\,\Big(\frac{\lambda}{(4\pi)^2}\Big)^{n+\frac{1}{2}}\,I_{2n+1}(\sqrt{\lambda})\, .
\eea

\noindent Thus

\begin{equation}
\langle O_n^{\mathbb{R}^4}W\rangle_ {\mathbb{S}^4}=n\,\Big(\frac{\lambda}{(4\pi)^2}\Big)^{\frac{n}{2}}\,I_n(\sqrt{\lambda})\, ,
\end{equation}
which is in perfect agreement with \cite{Giombi:2009ds}, yet through a completely different matrix model 
--see eq.(4.21) in  \cite{Giombi:2009ds} for $A_1=A_2=\frac{A}{2}=2\pi$.

\section{Correlators for chiral primary operators and Wilson loops in $\mathcal{N}=2$ superconformal SQCD}\label{SQCD}

We can extend our computations of correlators between Wilson loops and CPO's to the case of other $\mathcal{N}=2$ theories. 
In perturbation theory, one can express correlators in any  $\mathcal{N}=2$ superconformal field theory in terms of correlators of the  $\mathcal{N}=4$ theory.
As an illustration, we consider the   example  of $\mathcal{N}=2$ superconformal SCQD. For this case, the map between $\mathbb{R}^4$ and  $\mathbb{S}^4$ operators was partially worked out in \cite{Rodriguez-Gomez:2016ijh} at weak coupling. 
In the case of $\mathcal{N}=2$ superconformal SQCD, the map  $O^{\mathbb{R}^4}\rightarrow O^{\mathbb{S}^4}$ will be given through a power-series correction to the Chebyshev polynomial prescription of $\mathcal{N}=4$ theory.
 This can be traced back to the fact that the partition function for $\mathcal{N}=2$ superconformal QCD is (we assume that the gauge group is $U(N)$ for definiteness)

\begin{equation}
\label{matrixmodelZN2}
\mathcal{Z}_{\mathcal{N}=2} =\int d^N a\ \Delta(a) \frac{\prod_{i<j}H(a_i-a_j)^2}{\prod_i H(a_i)^{2N}}\,|e^{-2\pi {\rm Im}\tau_{YM}\sum a_i^2}  | \ \mathcal{Z}_{\rm inst}\, ,
\end{equation}
$$
 H(x)=\prod_{n=1}^{\infty}\Big(1+\frac{x^2}{n^2}\Big)^{n^2}e^{-\frac{x^2}{n}}\, .
$$
We shall consider the perturbative expansion in the zero instanton number sector, so we set 
$\mathcal{Z}_{\rm inst}\to 1$. A perturbative series is obtained by expanding the one-loop factor in powers of $a_i$. 
We use

\begin{equation}
\ln H(x)= -\sum_{n=2}^\infty (-1)^n\frac{\zeta(2n-1)}{n}x^{2n}\, .
\end{equation}
Now we can expand $\mathcal{Z}_{\mathcal{N}=2}$ as

\begin{eqnarray}
\label{e4}
\mathcal{Z}&=&\mathcal{Z}_{\mathcal{N}=4}\Big\{1-\zeta(3)\Big(3\langle {\rm Tr}\phi^2{\rm Tr}\overline{\phi}^2\rangle_{S^4}^{\mathcal{N}=4}-4\langle {\rm Tr}\phi^3{\rm Tr}\overline{\phi}\rangle_{S^4}^{\mathcal{N}=4}\Big)\\ && \nonumber  -\frac{2}{3}\zeta(5)\Big(10\langle {\rm Tr}\phi^3{\rm Tr}\overline{\phi}^3\rangle_{S^4}^{\mathcal{N}=4}-15 \langle {\rm Tr}\phi^4{\rm Tr}\overline{\phi}^2\rangle_{S^4}^{\mathcal{N}=4}+6\langle {\rm Tr}\phi^5{\rm Tr}\overline{\phi}^1\rangle_{S^4}^{\mathcal{N}=4}\Big)+\cdots\Big\}\, ;
\end{eqnarray}
where $\mathcal{Z}_{\mathcal{N}=4}$ is the $U(N)$ $\mathcal{N}=4$ SYM partition function and $\langle {\rm Tr}\phi^n{\rm Tr}\overline{\phi}^m\rangle_{S^4}^{\mathcal{N}=4}$ refers to the 2-point function of the ${\rm Tr}\phi^n$, ${\rm Tr}\overline{\phi}^m$ operators in the $\mathcal{N}=4$ SYM matrix model on the $S^4$. In this way we write the partition function for $\mathcal{N}=2$ superconformal QCD   solely in terms of quantities in $\mathcal{N}=4$ SYM at arbitrary $N$. In fact, using the results in the previous sections one can see that, at large $N$

\begin{equation}
\label{ZN2UN}
\mathcal{Z}_{\mathcal{N}=2}=\mathcal{Z}_{\mathcal{N}=4}\Big\{ 1 - \frac{3N^4\,\zeta(3)}{16\pi^2 {\rm Im}\tau_{YM}^2} +\frac{30N^5\zeta(5)}{96\pi^3 {\rm Im}\tau_{YM}^3}+\cdots\Big\}\, .
\end{equation}
Since, for instance, the correlator $\langle O_2^{\mathbb{R}^4}\overline{O}_2^{\mathbb{R}^4}\rangle_{\mathbb{R}^4}=\langle O_2^{\mathbb{S}^4}\overline{O}_2^{\mathbb{S}^4}\rangle_{\mathbb{S}^4}\sim \partial_{\tau_{YM}}\partial_{\overline{\tau}_{YM}}\ln \mathcal{Z}_{\mathcal{N}=2}$, we see that the correlator in $\mathcal{N}=2$ superconformal SQCD can be written in terms of correlators in $\mathcal{N}=4$. It is clear that, \textit{mutatis mutandi}, this result extends to all other correlators in $\mathcal{N}=2$ superconformal QCD.

Let us now turn to the correlator between CPO's and Wilson loops. For simplicity, we will focus on the correlator between $O_2^{\mathbb{R}^4}={\rm Tr}\phi^2$ and the Wilson loop. As shown in \cite{Rodriguez-Gomez:2016ijh}, $O_2^{\mathbb{R}^4}\rightarrow O_2^ {\mathbb{S}^4}$. Moreover, the correlators between $O_2^ {\mathbb{S}^4}$ and any other even-dimensional operators can be read off from the formula

\begin{eqnarray}
\label{dospuntos}
\langle O_{2n}^ {\mathbb{S}^4}\overline{O}_{2m}^ {\mathbb{S}^4}\rangle_ {\mathbb{S}^4}&=&\Big(\frac{\lambda}{4\pi^2}\Big)^{m+n}\frac{\Gamma(m+\frac{1}{2})\Gamma(n+\frac{1}{2})}{\pi(m+n)\Gamma(n)\Gamma(m)}\\ \nonumber &&-\frac{3}{4}\zeta(3)\Big(\frac{\lambda}{4\pi^2}\Big)^{m+n+2}\frac{(3+m+n+mn)\Gamma(m+\frac{1}{2})\Gamma(n+\frac{1}{2})}{\pi(m+1)(n+1)\Gamma(m)\Gamma(n)}+\cdots \, ;
\end{eqnarray}
where the first line corresponds to the leading order term, which is identical to that of the free field theory (and consequently analogous to that of $\mathcal{N}=4$ SYM due to the non-renormalization theorem) and the second represents the NLO correction.  Correlators with odd dimensional operators vanish.

Consider the correlator
\be
\langle O_2^{\mathbb{R}^4}W\rangle_ {\mathbb{S}^4}^{\rm SQCD} 
= \sum_{r=0}^{\infty}\frac{(2\pi)^{2r}}{(2r)!} \langle O_2^{\mathbb{S}^4} \overline{O}_{2r}^ {\mathbb{S}^4}\rangle_ {\mathbb{S}^4}\ .
\ee
Substituting the 2-point function (\ref{dospuntos}), we find
\begin{eqnarray}
\langle O_2^{\mathbb{R}^4}W\rangle_ {\mathbb{S}^4}^{\rm SQCD} &=& 
\frac{\lambda}{8\pi^2}\sum_{r=1}^\infty  \frac{1}{(r-1)!(r+1)!} \frac{\lambda^r}{4^r} -
\frac{3 \zeta(3)\lambda^3}{512\pi^6} \sum_{r=1}^\infty  \frac{(2+r)}{(r-1)!(r+1)!} \frac{\lambda^r}{4^r}
\nonumber\\
&=&\frac{\lambda}{8\pi^2}\,I_2(\sqrt{\lambda})\Big(1-\frac{9\zeta(3)}{64\pi^4}\lambda^2\Big)-\frac{3 \zeta(3)}{1024\pi^6} \lambda^{\frac{7}{2}}I_3(\sqrt{\lambda})+\cdots 
\nonumber \\ && =\frac{\lambda^2}{64\pi^2}+\frac{\lambda^3}{768\pi^2}+\frac{\lambda^4}{24576\pi^2}(1-\frac{54 \zeta(3)}{\pi^4})+\cdots\, .
\end{eqnarray}
By the same procedure, one may compute the analogous correlators for CPO's of higher dimensions using the explicit maps given in  \cite{Rodriguez-Gomez:2016ijh}.

\subsection{The $SU(N)$ case}

As emphasized above, \eqref{e4} is valid for any gauge group. Let us now particularize it and compute the partition function for $\mathcal{N}=2$ SQCD with $SU(N)$ gauge group at finite $N$. Imposing that ${\rm Tr}\phi=0$, we get

\begin{equation}
\mathcal{Z}_{\mathcal{N}=2}^{SU(N)}=\mathcal{Z}^{SU(N)}_{\mathcal{N}=4}\Big\{1-3\zeta(3)\langle {\rm Tr}\phi^2{\rm Tr}\overline{\phi}^2\rangle_{S^4}^{\mathcal{N}=4,\,SU(N)}+\cdots\Big\}\, .
\end{equation}
It is useful now to re-write $\langle {\rm Tr}\phi^2{\rm Tr}\overline{\phi}^2\rangle_{S^4}^{\mathcal{N}=4,\,SU(N)}$ in terms of VEV-less operators $O_2$ in the $SU(N)$ $\mathcal{N}=4$ theory as

\begin{equation}
\langle {\rm Tr}\phi^2{\rm Tr}\overline{\phi}^2\rangle_{S^4}^{\mathcal{N}=4,\,SU(N)}=\langle O_2\overline{O}_2\rangle_{S^4}^{\mathcal{N}=4,\,SU(N)}+V_2^2\, ,
\end{equation}
being $V_2$ the VEV of ${\rm Tr}\phi^2$ in the $SU(N)$ theory. Note that both $V_2$ and $\langle {\rm Tr}\phi^2{\rm Tr}\overline{\phi}^2\rangle_{S^4}^{\mathcal{N}=4,\,SU(N)}$ follow from the partition function $\mathcal{Z}_{\mathcal{N}=4}^{SU(N)}$ through derivatives. Recall that (see \textit{e.g.} \cite{Drukker:2000rr})

\begin{equation}
\label{ZN4}
\mathcal{Z}_{\mathcal{N}=4}^{SU(N)} =\sqrt{2\,N\,{\rm Im}\tau_{YM}}\, \mathcal{Z}_{\mathcal{N}=4}^{U(N)}\, ;\qquad \mathcal{Z}^{U(N)}_{\mathcal{N}=4}=\Big(4\pi{\rm Im}\tau_{YM}\Big)^{-\frac{N^2}{2}}\, \,(2\pi)^{\frac{N}{2}}\, G(N+2)\, .
\end{equation}
Therefore

\begin{equation}
\langle O_2\overline{O}_2\rangle_{S^4}^{\mathcal{N}=4,\,SU(N)}=\pi^{-2}\partial_{\tau_{YM}}\partial_{\overline{\tau}_{YM}}\ln \mathcal{Z}_{\mathcal{N}=4}^{SU(N)}=\frac{N^2-1}{8\pi^2 {\rm Im}\tau_{YM}^2}\, ,
\end{equation}
and

\begin{equation}
V_2=i\pi^{-1}\partial_{\tau_{YM}}\ln \mathcal{Z}_{S^4}^{\mathcal{N}=4,\,SU(N)}=\frac{N^2-1}{4\pi {\rm Im}\tau_{YM}}\, .
\end{equation}
With this at hand

\begin{equation}
\label{ZN2SUN}
\mathcal{Z}_{\mathcal{N}=2}^{SU(N)}=\mathcal{Z}_{\mathcal{N}=4}^{SU(N)}\Big\{ 1-\frac{3\zeta(3)\,(N^4-1)}{16\pi^2{\rm Im}\tau_{YM}^2}+\cdots \Big\}\ .
\end{equation}

Let us now consider correlators for CPO's. The special case of correlators for ${\rm Tr}\phi^2$ is  simple, as these insertions arise from derivatives of the YM coupling $\tau_{YM}$.
 The case of the $SU(N)$ theory is particularly interesting, since a conjecture for $\langle{\rm Tr}\phi^2{\rm Tr}\overline{\phi}^2\rangle_{\mathbb{R}^4}$ was put forward in \cite{Rodriguez-Gomez:2016ijh}. Using \eqref{ZN2SUN} one finds

 \begin{equation} \label{cofi}
\langle {\rm Tr}\phi^2{\rm Tr}\overline{\phi}^2\rangle^{\mathcal{N}=2,\,SU(N)}_{\mathbb{R}^4}=\frac{2(N^2-1)}{\pi^2{\rm Im}\tau_{YM}^2}-\frac{9\zeta(3)\,(N^2-1)(N^2+1)}{2\pi^4{\rm Im}\tau_{YM}^4}+\cdots\, ,
\end{equation}
which demonstrates the conjecture of \cite{Rodriguez-Gomez:2016ijh}.
In the particular case of $SU(2)$ gauge group,  this formula reproduces  an earlier result given in \cite{Gerchkovitz:2016gxx} and, for  $SU(3),\ SU(4)$,
the formula (\ref{cofi}) also reproduces the expressions given in \cite{Baggio:2015vxa} (see also \cite{Gerchkovitz:2016gxx}).

\section{Conclusions}\label{conclusions}

In this paper we have analyzed the  structure of the operator mixing arising when mapping the $\mathbb{R}^4$ theory into the $\mathbb{S}^4$. Concentrating on the $\mathcal{N}=4$ SYM case, and at large $N$, such operator mixings are elegantly  encoded into Chebyshev polynomials. This provides a general expression for the mixing coefficients, whose $\partial_{\tau_{\rm YM}}\partial_{\overline{\rm \tau}_{\rm YM}}$  derivative is an unambiguous quantity $A_n^{(2k)}$.
It would be  be extremely interesting to understand if the gauge-invariant mixing coefficients $A_n^{(2k)}$ 
can be  computed in the holographic dual setup for supergravity in $AdS_5\times S^5$ space.

We have provided very non-trivial evidence of the relevant role of the Chebyshev polynomial mixing by computing the $\langle WO_n^{\mathbb{R}^4}\rangle_{\mathbb{R}^4}$ correlator. 
In \cite{Giombi:2009ds}, such correlators were computed by means of a completely different matrix model --namely the two-matrix model. Here we have reproduced the result \cite{Giombi:2009ds} for the correlation function $\langle W O_n^{\mathbb{R}^4} \rangle_{\mathbb{R}^4} $ from first-principles, using the localized partition function and disentangling the operator mixing induced by mapping into $\mathbb{S}^4$.
Thus, our result proves the conjectural relation proposed in \cite{Giombi:2009ds} with the two-matrix model.
In addition, it suggests a novel duality between two completely different 
matrix models at large $N$. It would be very interesting to explore such ``duality" and understand its origins, in particular under the light of the connection of the two-matrix model to 2d Yang-Mills, 
and the extent to which this duality holds. In particular, it would be very interesting to understand the generic correlators $\langle W\prod_i O_{n_i}^{\mathbb{R}^4}\rangle_{\mathbb{R}^4}$. Note that the Chebyshev polynomial prescription 
does not take into account mixing with multitrace operators in the $\mathbb{S}^4$, which, on the other hand, might be relevant whenever there is more than one CPO in the correlator with the Wilson loop. Understanding this point would be very interesting. 

We have also computed the  $\langle WO_n^{\mathbb{R}^4}\rangle_{\mathbb{R}^4}$ correlator
in  $\mathcal{N}=2$ superconformal SQCD up to (and including) $O(\lambda^4)$ in perturbation theory. Along the way, we have seen that correlators of CPO's in $\mathcal{N}=2$ superconformal QCD can be written in terms of correlators of CPO's in $\mathcal{N}=4$ SYM. In particular, this allowed us to provide a simple proof of 
the  formula for the finite $N$ 
two-point function of the $O_2^{\mathbb{R}^4}$ operator conjectured in \cite{Rodriguez-Gomez:2016ijh}. Further studying this connection among CPO correlators of $\mathcal{N}=2$ SQCD and $\mathcal{N}=4$ SYM would be very interesting.

\section*{Acknowledgements}

We thank M. Tierz for a useful discussion.
D.R-G si partly supported by the Ramon y Cajal grant RYC-2011-07593, the asturian grant FC-15-GRUPIN14-108, the spanish national grant MINECO-16-FPA2015-63667-P as well as the EU CIG grant UE-14-GT5LD2013-618459. D.R-G. would like to thank the U. Barcelona for warm hospitality during the initial stages of this project. J.G.R. acknowledges financial support from projects  FPA2013-46570, 2014-SGR-104 and  MDM-2014-0369 of ICCUB (Unidad de Excelencia `Mar\'ia de Maeztu').

\end{document}